\documentclass[
	a4paper, %
	10pt, %
	unnumberedsections, %
	twoside, %
]{LTJournalArticle}

\runninghead{Clicks Versus Conversion: Choosing a Recommender's Training Objective in E-Commerce}

\usepackage{color,soul}
\usepackage{tabularx}
\usepackage{booktabs}
\usepackage{enumitem}
\setlist[itemize]{leftmargin=*}

\AtBeginDocument{%
  }

\author{%
	Michael Weiss\textsuperscript{1}, Robert Rosenbach\textsuperscript{1} and Christian Eggenberger\textsuperscript{1}\thanks{Corresponding author: \href{mailto:michael.weiss@digitecgalaxus.ch}{michael.weiss@digitecgalaxus.ch}}
}

\begin{document}

\title{Clicks Versus Conversion:\\ Choosing a Recommender's Training Objective in E-Commerce}

\date{\footnotesize\textsuperscript{\textbf{1}}Digitec Galaxus AG, Zurich, Switzerland}

\renewcommand{\maketitlehookd}{%
    \begin{abstract}
        Ranking product recommendations to optimize for a high click-through rate (CTR) or for high conversion, such as add-to-cart rate (ACR) and Order-Submit-Rate (OSR, view-to-purchase conversion) are standard practices in e-commerce.
        Optimizing for CTR appears like a straightforward choice: Training data (i.e., click data) are simple to collect and often available in large quantities. 
        Additionally, CTR is used far beyond e-commerce, making it a generalist, easily implemented option.
        ACR and OSR, on the other hand, are more directly linked to a shop's business goals, such as the Gross Merchandise Value (GMV). 
        In this paper, we compare the effects of using either of these objectives using an online A/B test.
        Among our key findings, we demonstrate that in our shops, optimizing for OSR produces a GMV uplift more than five times larger than when optimizing for CTR, without sacrificing new product discovery. Our results also provide insights into the different feature importances for each of the objectives.
    \end{abstract}
}

\maketitle

\section{Introduction}

Product recommendations are prevalent in modern e-commerce, helping users better discover the available range of products and, by doing so, improving the shop's business metrics (revenue, profit, etc.).
Recommendations are typically \emph{ranked} to optimize one or multiple desired objectives~\cite{jannach2022multi}, such as the click-through-rate (CTR), the Add-to-Cart-Rate (ACR) or the view-to-purchase conversion (Order-Submit-Rate, OSR)~\cite{stalidis2023recommendation}. This is done automatically using a ranking model. The choice of the optimization target is not always obvious: Prioritizing CTR products may increase the engagement with the store, but at the same time also give a higher priority to clickbait products with less actual relevance to the customer. Conversion, on the other hand, can be hard to optimize for as there are naturally much fewer training samples available than for the CTR~\cite{tsagkias2021challenges}, which may hinder the model's generalization capability to less frequently bought products and product categories. 

To the best of our knowledge, the majority of related literature primarily considers click events when discussing ranking models, thus optimizing for CTR. This choice is natural as among other reasons click events are applicable pretty much in every domain, and as clicks naturally occur more often than subsequent events (as each subsequent event implies a previous click). However, the implied correlation between CTR and business metrics may not be as strong as often assumed~\cite{tsagkias2021challenges}. 

In this paper, we empirically investigate the implications that come from the choice of optimization target when training a ranking model in an e-commerce setting. 
Specifically, we aim to shed light on the trade-off between choosing CTR, ACR and OSR as prediction target during training.
In addition, we complement existing works on the importance of specific input features~\cite{jannach2017session} by comparing the most influential inputs for each optimization objective.

We run our study as online A/B/C/D-Test (an A/B Test with four test groups) on our DigitecGalaxus shops (mainly \href{https://galaxus.ch}{galaxus.ch}, \href{https://digitec.ch}{digitec.ch} and~\href{https://galaxus.de}{galaxus.de}), which not only represent the leading B2C e-commerce platforms in Switzerland with millions of active customers yearly, but are also a major player in the EU market. Our product catalog includes more than 25 million products from different sectors including, e.g. IT, Supermarket, Fashion, and Furniture. In sum, this thus provides a broad and international environment for our study.

\section{Experimental Setup}

We illustrate the impact of selecting an appropriate optimization objective using the example of the "similar items" recommender on our page. Below we will briefly introduce this recommender, its relevant terminology, and the test setup.

\subsection{Used Recommendation System}

\begin{figure*}[t]
  \centering
  \includegraphics[width=.7\linewidth]{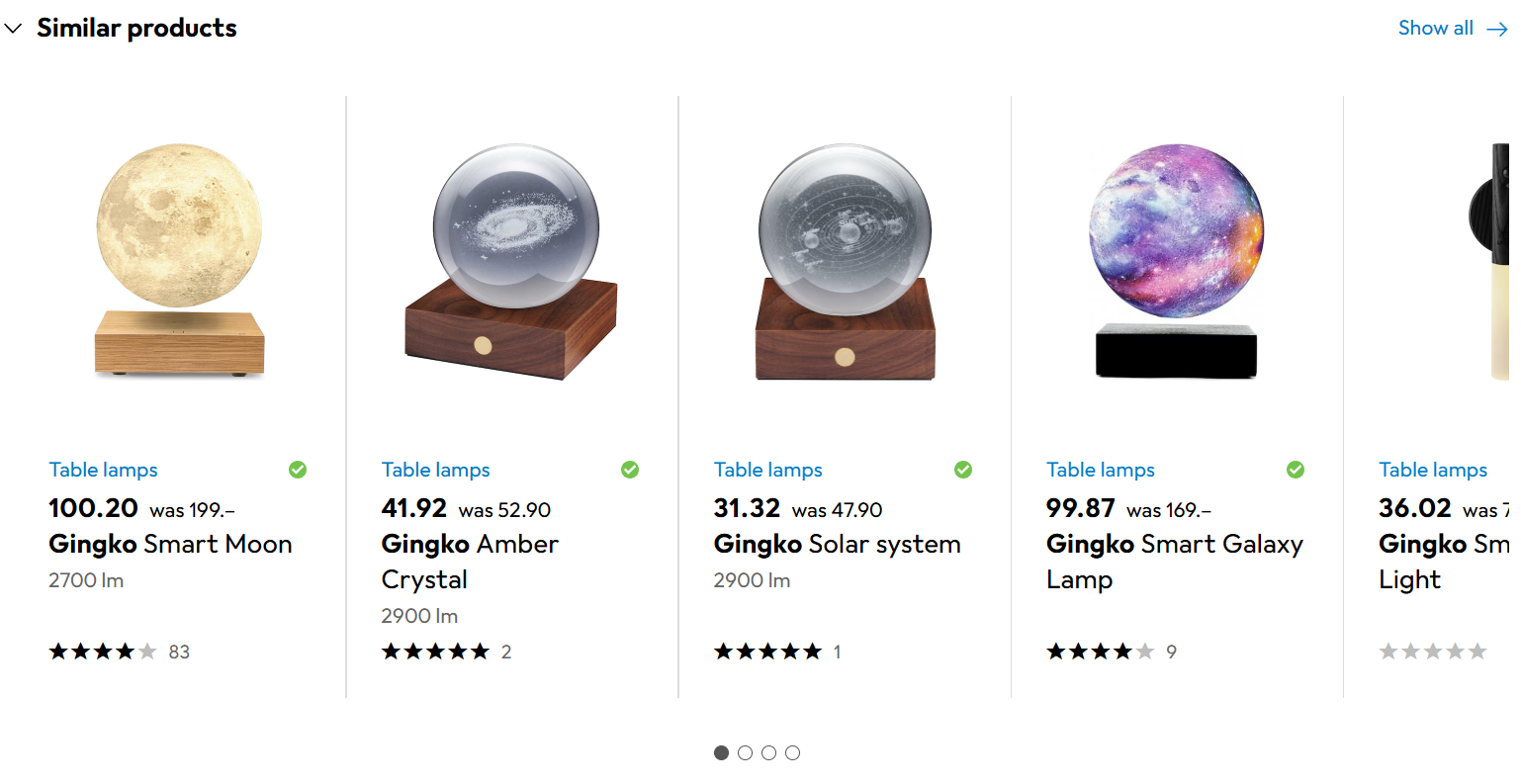
}
  \caption{\emph{Similar Products} for \href{https://www.galaxus.ch/en/s14/product/gingko-smart-moon-table-lamps-16457822}{"Gingko Smart Moon" lamp}}
  \label{fig:similar_prod}
\end{figure*}

The test was conducted on the "similar items" recommender system. This recommender system helps users on a product page--the designated page for a specific product--to explore alternatives to this product. \autoref{fig:similar_prod} provides an example of said recommender as shown to our users.

The recommender shows up to 16 similar items. When the recommender enters the user's viewport, we record a block impression. Each block impression contains up to 16 tile impressions—one for each product that appears in the user's viewport. Opened on a standard desktop computer, four and a half of these are in the viewport by default (shown in \autoref{fig:similar_prod}).

Our recommender system follows a standard four-stage architecture~\cite{higley2022building}, consisting of Retrieval, Filtering, Scoring and Ranking. %
Specifically, products are retrieved based on their behavioral similarity, i.e., the similarity in user visits which is calculated using a simple Word2Vec model~\cite{mikolov2013word2vec}, with products representing words and sessions representing sentences, similar to Meta-Prod2Vec~\cite{vasile2016meta}.
Products not meeting some strict requirements (e.g. availability) are then filtered out.
Scoring models are then used to predict a product's \emph{success rate}, i.e., its click-, add-to-cart-, or order-submit probability, respectively. This is explained in detail in the next section.
The orderer then sorts the products by their score. Lastly, it also applies some minor re-orderings to account for superior requirements (e.g. marketing related constraints).

\subsection{Scoring Models}
\begin{table}[b]
    \centering
    \small
    \begin{tabular}{lccc}
    \toprule
                       & \textbf{click}        &\textbf{add-to-cart}    &\textbf{order-submit} \\
    \midrule
    \textbf{total samples}      & 3'000'000   & 3'000'000      & 3'000'000   \\
    \textbf{negative class} & 2'583'908   & 2'549'426      & 2'529'132   \\
    \textbf{positive class}     &   416'092   &   450'574      &   470'868   \\
\bottomrule
    \end{tabular}
    \caption{Data set sizes used for model training}
    \label{tab:dataset}
\end{table}

To determine the item ordering in the recommender system's third step we use point-wise ranking. This is done via an XGBoost model~\cite{chen2016xgboost}, which predicts a score for the target objective, which is either \emph{click}, \emph{add-to-cart} or \emph{order-submit}.

We extracted the training datasets for all three models from the same base dataset. It consists of 12 weeks of block impressions from the first quarter of 2025 and includes identical features--such as product price, ratings, traffic, sales, etc.--for each model. This base data set is adjusted for each of the optimization goals by using all block-impressions where at least one tile-impression had a success-event (as defined by the target variable). 
To ensure fair comparison, we created equally sized training datasets for all variants by sampling block impressions until each dataset contained three million tile impressions. The skew of the three resulting data sets is between 13\% and 16\% for all models, as shown in \autoref{tab:dataset}.

Given these data sets, each model with its data set is optimized via a standard grid search with identical parameter ranges. The best model for each optimization target is then used to perform a point-wise ranking of recommended items according to its optimization goal, and the recommendations are ordered by these predicted scores in descending order.

\section{Results}

\begin{figure*}[t]
  \centering
  \includegraphics[width=0.6\linewidth]{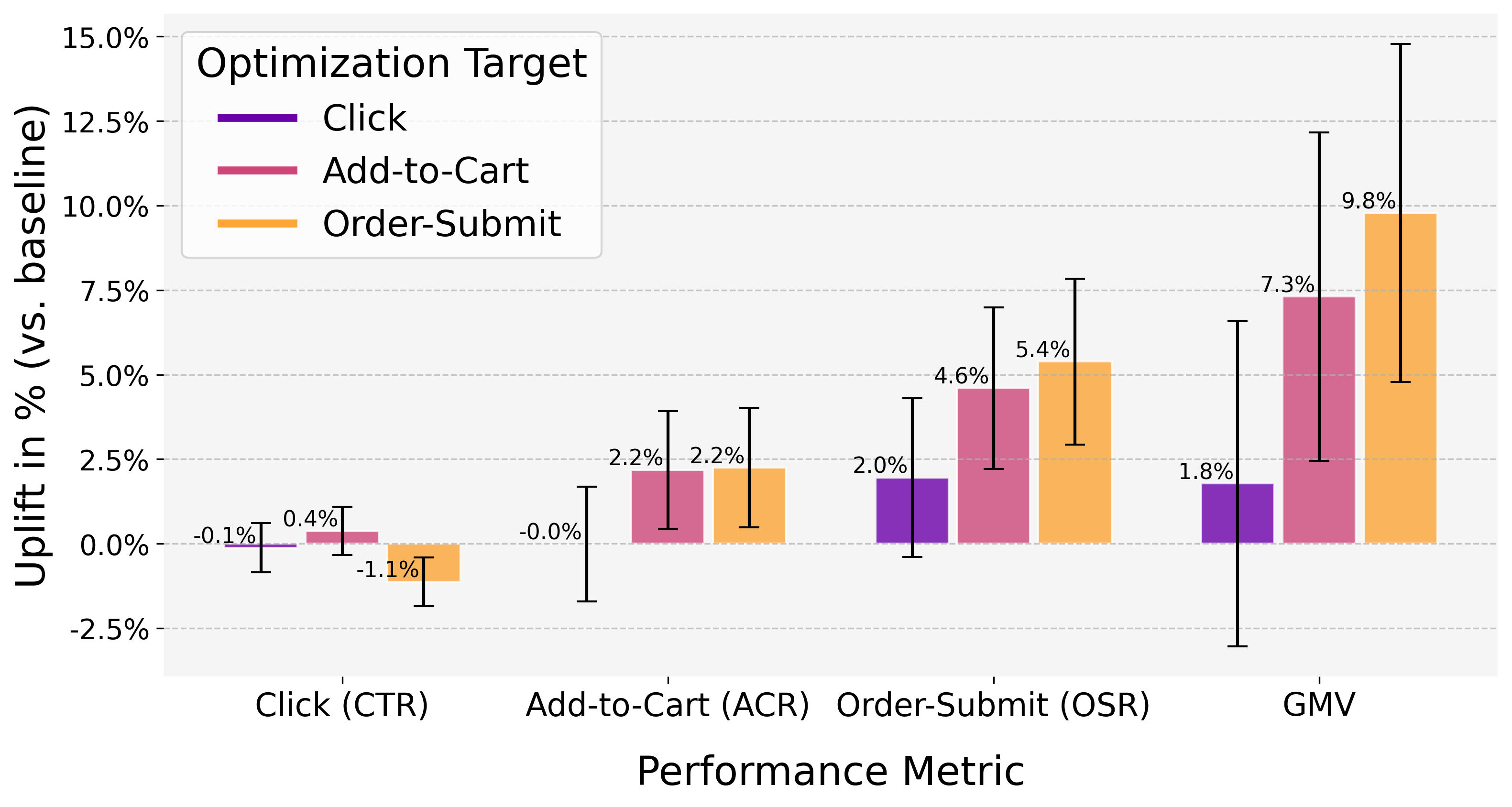}
  \caption[]{Observed uplift of the scoring models for the three optimization targets in the e-commerce funnel\footnotemark}
  \label{fig:results}
\end{figure*}

We compared the effects of the three models as an A/B/C/D test. Sorting by the session-based product similarity used for retrieval, without any dedicated ranking model, was used as the baseline variant.
The test ran for a duration of twelve days, starting on April 30, 2025, resulting in about 1.8 million block impressions per test group. 

In the following, we first report the results of our primary research question - the influence of the optimization objective - and subsequently present additional insights we observed when running the experiment.

\subsection{Optimization Objective}

\footnotetext{Error bars represent the 95\%-Confidence-Intervals, with standard errors clustered at the randomization unit (i.e., the device-browser combination). To limit the influence of outliers, GMV values are capped at the 99.9th percentile.}

In evaluating the A/B test, we are mostly interested in click- and post-click performance. We track whether an item presented to the user was clicked, whether the clicked item was subsequently added to the cart, and whether it was eventually purchased. This progression throughout the e-commerce funnel lets us compare the different versions of the recommender system in detail. Obviously these metrics need to be balanced with additional health metrics (to cover additional important aspects such as novelty), but they serve well as first-line performance evaluation.

Our results are shown in \autoref{fig:results}. While the models optimized to improve add-to-cart and order-submit achieve a significant uplift in ACR, OSR and GMV (per block impression), this is not true for the model optimized to improve CTR. In particular, the model optimized to improve CTR does not achieve its goal. This reflects the strong baseline performance of our session-based customer behavior similarity model, which inherently captures click patterns.%

The model aimed at optimizing order-submits does so at the expense of clicks: A minor, but significant, decrease in CTR is more than overcompensated by the achieved increases in OSR, and ultimately provides the best optimization in terms of business objectives. 
The superiorities of the ACR and OSR models over the CTR model are significant w.r.t the observed ACR, OSR and GMV ($p<0.05$), highlighting the insufficiency of CTR optimization when aiming for revenue.

It is worth noting that results of previous A/B tests are consistent with the main results presented in this paper, and over different countries and the vast majority of product segments. This consistency across multiple tests also strengthens our confidence.%

\begin{table*}[t]
    \centering
    \small
    \begin{tabularx}{\textwidth}{lXXX}
    \hline
     & \textbf{click}    & \textbf{add-to-cart}    & \textbf{order-submit} \\
    \toprule
    \textbf{Feature \#1}  & behavioral similarity      & behavioral similarity   & behavioral similarity \\
    \textbf{Feature \#2}  & rating upper confidence interval & visual similarity       & item seen previously \\
    \textbf{Feature \#3}  & rating average             & item seen previously    & item sales rank (last 28d) \\
    \textbf{Feature \#4}  & item traffic percentile    & relative item sales rank (last 28d)    & visual similarity \\
   \textbf{Feature \#5}  & visual similarity          & item sales total & item sales fraction in product type (last 28d) \\
    \bottomrule
    \end{tabularx}
    \caption{Most important features per optimization goal}
    \label{tab:modelfeatures}
\end{table*}

\subsection{Further Insights}
Our study allowed us to make a range of secondary observations that might be of interest to the reader. Due to space constraints, we only mention them briefly:
\begin{itemize}
    \item We observe an inherently different predictiveness w.r.t. distinct optimization goals, which translates to distinct uplift patterns if re-ranking based on the prediction scores is performed.
    \item We see a clear distinction in feature uptake of the single models. 
    As shown in \autoref{tab:modelfeatures}, while user behavior is the most important feature in any case, the most relevant features for the CTR model include rating-focused and traffic features, whereas the ACR and OSR models rely more on sales-based features and previous item interaction. This complements existing work~\cite{jannach2017session}.
    \item Although only the ACR and OSR models rely heavily on features describing previous interactions, all variants show a similar \emph{discovery share} (i.e., percentage of clicks that lead to users' seeing a product for the first time within their 14-day browsing history). 
    \item We observed different time to saturation on the distinct models' learning curves, with the CTR model converging first, followed by the ACR and OSR models. This may indicate that the better availability of click data does not necessarily pose an advantage for systems at scale, as the models learning curve already saturated.
\end{itemize}

\subsection{Discussion}
Our experiments highlight the relevance of choosing the optimal prediction target when training a recommender ranking model. By switching just the training objective, all other things equal, our primary business value (measured by GMV) was significantly influenced. 
Since this change requires modifying only a few lines of code, testing different optimization targets represents a low-hanging fruit for improving recommender system performance. 
All three models maintain comparable discovery shares, alleviating concerns that optimizing for revenue might sacrifice the exploration benefits typically associated with CTR optimization~\cite{Falk2023Behind}.

Our results are particularly interesting w.r.t. CTR, which may be considered a "default" optimization target by many, and described to be closely related to conversion ~\cite{zhang2023dive}.
In practice, our results indicate that it is worth sacrificing some CTR, even compared to the baseline without a dedicated ranking model, to achieve an overall more favorable result. This also provides important context to related literature which mentions simultaneous optimization of CTR and OSR~\cite{bagherjeiran2010ranking}.
These complementarities are also of paramount importance when building training objectives based on multiple objectives.~\cite{jannach2022multi, li2021attentive}

However, most importantly, this finding is crucial for practitioners who implement an e-commerce recommendation system. Our results show that when aiming to optimize revenue, the correlation between CTR and OSR is not generally strong enough to allow optimizing for the first, even though CTR optimization may be somewhat easier to implement, as click data is easily tracked and available in high quantities.

\section{Conclusion}
Our results allow multiple interesting observations regarding training objectives, and quantifies them with real-world data. Although the findings are to some extent expected, our results emphasize the importance of such considerations in real-world settings. The insights presented on the impact of different optimization objectives are of high practical relevance, and may also provide interesting real-world motivation for future theoretical work, such as the study of multi-objective recommender systems~\cite{jannach2022multi}.

\subsection{Threats to Validity}
Our experiments are limited to one company, and even though this includes multiple stores, they all share a very similar layout and design, which may cause bias. Our study is also limited to about two weeks, thus not taking seasonality into account. As such, we do not claim generalizability of the specific observed results. Instead, we hope that our paper encourages others to run similar experiments and to identify optimal training objectives for their recommenders.

As described above, we made sure that the training datasets for all variants have the same size, for fair comparison. Especially for small shops with constrained data availability, the higher availability of click data than data points collected later in the e-commerce funnel will need to influence the choice of optimization target as well. As for our shops, we have seen in previous A/B tests that more click data does not improve the CTR model any further, which is in line with the above described observation of the CTR models quick learning curve saturation.

\printbibliography

\newpage
\section*{Author Biographies}

\subsection*{Michael Weiss}
Michael Weiss is a Senior Software Engineer at DigitecGalaxus, primarily responsible for the low-latency, reliable and safe online integration of the recommenders systems machine learning components. He holds a PhD in Informatics from the Università della Svizzera italiana. His previous work mostly focuses around the fail-safe integration of machine learning components by means of uncertainty quantification and outlier detection. 

\subsection*{Robert Rosenbach}
Robert Rosenbach is a senior data scientist at DigitecGalaxus, specializing in the design, development, and evaluation of machine learning algorithms for recommender systems, with a strong focus on measurable business impact.
He holds a PhD in theoretical physics from the University of Ulm, where his research focused on simulating open quantum systems using machine learning techniques.

\subsection*{Christian Eggenberger}
Christian Eggenberger is a Senior Data Analyst at DigitecGalaxus, primarily responsible for the development of the experimentation framework at DigitecGalaxus. He supports the recommendations domain with causal impact analyses. He holds a PhD in Management and Economics from the University of Zurich.

\end{document}